\documentclass[]{aastex}

\usepackage{emulateapj5}
\usepackage{onecolfloat}
\usepackage{graphicx}
\usepackage{fancyheadings}
\usepackage{ulem}
\usepackage{rotating}
\usepackage{lscape}

\def\arcmin{\hbox{$^\prime$}}
\def\arcsec{\hbox{$^{\prime\prime}$}}
\def\arcdeg{\mbox{$^\circ$}}%
\def\fd{\hbox{$~\!\!^{\rm d}$}}
\def\fh{\hbox{$~\!\!^{\rm h}$}}
\def\fm{\hbox{$~\!\!^{\rm m}$}}
\def\fs{\hbox{$~\!\!^{\rm s}$}}

\newcommand{\cmjj}{\mbox{${\rm cm^{-2}}$}}
\newcommand{\etal}{et al.}
\newcommand{\hI}{\mbox{${\rm H\ I}$}}
\newcommand{\kms}{\mbox{km\ s${^{-1}}$}}
\newcommand{\lya}{\mbox{${\rm Ly}\alpha$}}
\newcommand{\lyb}{\mbox{${\rm Ly}\beta$}}

\begin{document}

\twocolumn[%
\lefthead{Chen \etal}
\righthead{}

\slugcomment{Accepted for Publication in the Astrophysical Journal Letters}

\title{ECHELLE SPECTROSCOPY OF A GRB AFTERGLOW AT $z=3.969$: A NEW PROBE OF THE
INTERSTELLAR AND INTERGALACTIC MEDIA IN THE YOUNG UNIVERSE}
\author{
HSIAO-WEN CHEN\altaffilmark{1,5}, 
JASON X.\ PROCHASKA\altaffilmark{2}, 
JOSHUA S.\ BLOOM\altaffilmark{3}, and
IAN B.\ THOMPSON\altaffilmark{4}}


\begin{abstract}

  We present an echelle spectrum of the {\it Swift} GRB 050730, obtained four 
hours after the burst using the MIKE spectrograph on the Magellan Clay 
Telescope when the afterglow was at $R=17.7$.  The spectrum reveals a forest of
absorption features superimposed on a simple power-law shaped continuum, best 
described as $f_\nu(\lambda)\propto \lambda^{\alpha}$ with $\alpha =1.88\pm 
0.01$ over $\lambda=7000-9000$ \AA.  We identify the GRB host at $z_{\rm GRB}=
3.96855$ based on the hydrogen Lyman absorption series, narrow absorption lines
due to heavy ions such as O\,I, C\,II, Si\,II, S\,II, Ni\,II, Fe\,II, C\,IV, 
Si\,IV, and N\,V, and fine structure transitions such as O\,I$^*$, O\,I$^{**}$,
Si\,II$^*$, C\,II$^*$, and Fe\,II$^*$.  Together these transitions allow us to 
study the the properties of the interstellar medium (ISM) in the GRB host.  The
principal results are as follows.  (1) We estimate a neutral hydrogen column 
density of $\log\,N(\hI)=22.15\pm 0.05$ in the host.  (2) The associated metal 
lines exhibit multiple components over a velocity range of $\sim 80$ \kms, with
$>90$\% of the neutral gas confined in 20 \kms.  (3) Comparisons between 
different ionic transitions show that the host has little/no dust depletion and
has $1/100$ solar metallicity.  (4) The absorbing gas has much higher density 
than that of intervening damped \lya\ absorption (DLA) systems.  In addition, 
we report the identification of an intervening DLA system at $z_{\rm DLA}=
3.56439$ with $\log\,N(\hI)=20.3\pm 0.1$ and $< 5$\% solar metallicity, a Lyman
limit system at $z_{\rm LLS}=3.02209$ with $\log\,N(\hI)=19.9\pm 0.1$, a strong
Mg\,II absorber at $z_{\rm Mg\,II}=2.25313$, and a pair of Mg\,II absorbers at
$z_{\rm Mg\,II}=1.7731$, 57 \kms\ apart.  We demonstrate that rapid echelle 
spectroscopy of GRB afterglows helps to reveal a wealth of information in the 
ISM and the intergalactic medium along the sightline which, when followed 
up with late-time, deep imaging, will allow us to uncover a sample of distant 
galaxies with known ISM properties to constrain galaxy formation models.

\end{abstract}

\keywords{gamma rays: bursts---ISM: abundances---ISM: kinematics---intergalactic medium}
]
\altaffiltext{1}{MIT Kavli Institute for Astrophysics and Space Research, 
Cambridge, MA 02139-4307, {\tt hchen@space.mit.edu}}

\altaffiltext{2}{UCO/Lick Observatory; University of California, Santa
  Cruz, Santa Cruz, CA 95064, {\tt xavier@ucolick.org}}

\altaffiltext{3}{Department of Astronomy, 601 Campbell Hall, University of 
California, Berkeley, CA 94720 {\tt jbloom@astron.berkeley.edu}}

\altaffiltext{4}{Observatories of the Carnegie Institution of Washington, 813 
Santa Barbara Street, Pasadena, CA 91101, U.S.A., {\tt ian@ociw.edu}}

\altaffiltext{5}{Current address: Department of Astronomy and Astrophysics,
University of Chicago, IL 60637, {\tt hchen@oddjob.uchicago.edu}}


\section{INTRODUCTION}

\setcounter{footnote}{0}

  Various surveys designed to detect emission at optical, near-infrared, and 
sub-mm wavelengths have yielded large samples of galaxies at redshift $z>2$ 
(e.g.\ Steidel \etal\ 1999; Blain \etal\ 2002), but whether these galaxies are 
representative of the galaxy population at high redshifts and how they are 
related to the local population is not clear.  Damped \lya\ absorption (DLA) 
systems probe high-redshift gaseous clouds of neutral hydrogen column density 
$N(\hI)$ consistent with what is observed in the disks of nearby luminous 
galaxies (e.g.\ Wolfe \etal\ 2005).  They are selected uniformly with $N(\hI) 
\ge 2 \times 10^{20}$ \cmjj\  and represent a unique sample of distant galaxies
with known interstellar medium (ISM) properties from absorption-line studies 
(e.g.\ Pettini \etal\ 1999; Prochaska \etal\ 2003).  Indentifying the stellar 
counterpart of the DLAs has, however, been challenging because of the glare of 
background quasars (Colbert \& Malkan 2002; Le Brun \etal 1997; Rao \etal\ 
2003; Chen \& Lanzetta 2003).

  Long-duration gamma-ray bursts (GRBs) are believed to originate in the death 
of massive stars (e.g.\ Woosley 1993; Paczy\'nski 1998; Bloom \etal\ 2002; 
Stanek \etal\ 2003). Some bursts are followed by optical afterglows (e.g.\ 
Akerlof \etal\ 1999) that can briefly exceed the absolute brightness of any 
known quasar by orders of magnitude and serve as bright background sources for 
probing intervening gas along the line of sight.  Because of their transient 
nature, however, optical afterglows do not interfere with follow-up studies of 
absorbing galaxies close to the sightlines.  Early-time, high-resolution 
spectroscopy of GRB afterglows together with deep, late-time imaging of 
galaxies along the sightlines offers a novel means to uncover a sample of
high-redshift galaxies based on their absorption properties.  In the context of
heirarchical structure formation, GRB progenitors can form before massive black
holes, and therefore may be used to probe early universe, well into the age of 
reionization.

  Prompt localization of GRB afterglows is critical in order to take advantage
of their brief but extreme brightness for acquiring echelle spectroscopy.  It
has been difficult in the past to carry out rapid spectroscopy for the optical
transient (OT) because of time-consuming processes to localize the bursts.  
Despite an extensive effort, only a small number of GRBs have been 
spectroscopically identified at $z>2$ with low-to-moderate resolution spectra 
available\footnote{see http://www.mpe.mpg.de/\char'176jcg/grbgen.html for a 
complete list.} and two with echelle data available (Fiore \etal\ 2005).  
Together these data show that all GRB host galaxies have abundant neutral gas, 
and some have the largest $N(\hI)$ among all DLA systems (Jensen \etal\ 2001;
M{\o}ller \etal\ 2002; Castro \etal\ 2003; Jakobsson \etal\ 2004; 
Vreeswijk \etal\ 2004).

\begin{figure*}[th]
\begin{center}
\includegraphics[scale=0.6, angle=270]{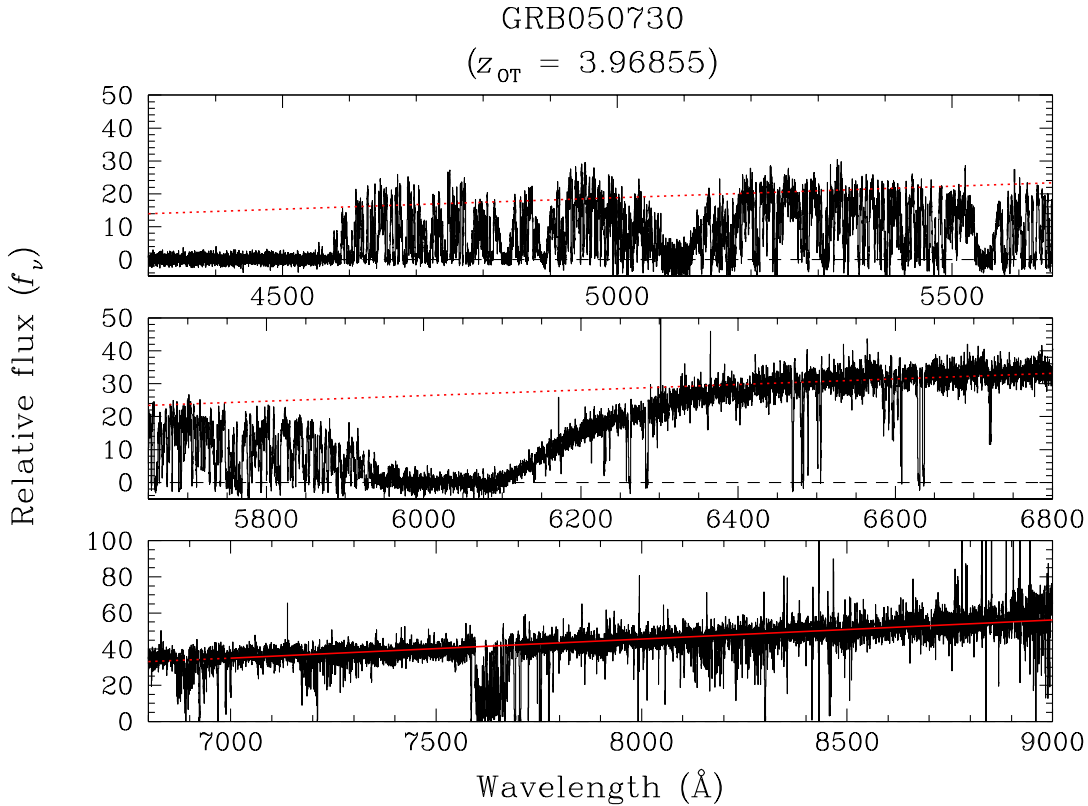}
\caption{Stacked echelle spectrum of the OT of GRB050730.  The signal-to-noise 
of the echelle data is $S/N=12$ at 4800 \AA\ and $S/N=13$ at 8000 \AA\ per 
resolution element.  Blueward of the \lya\ absorption trough at 6040 \AA\ from 
the GRB host environment is a forest of foreground \lya\ absorbers, including a
DLA system at 5550 \AA\ ($z_{\rm DLA}=3.564$).  The \lyb\ absorption feature of
the GRB host is apparent at 5090 \AA.  The absence of flux at wavelengths below
4530 \AA\ indicates little/no high energy photons beyond the Lyman limit 
transition escape the host of the GRB.  The continuum at $\lambda=7000-9000$ 
\AA\ is best-described by a power-law model $f_\nu(\lambda)\propto 
\lambda^{1.88}$ as shown in the solid *red) curve.  Note that the power-law fit
from 7000\AA\ to 9000\AA\ somewhat underpredicts the apparent continuum flux 
from 4600\AA\ to 5000\AA\ (dotted curve).}
\end{center}
\end{figure*}

  The current generation gamma-ray satellite, {\it Swift}, is designed to 
provide nearly-instant localization of new GRBs.  Over the past two years, 
our group has been pursuing well-localized GRB afterglows with moderate-to-high
resolution spectroscopy (e.g.\ Prochaska \etal\ 2004; Bloom \etal\ 2005).  The
primary goal of our project is to collect a statistically significant sample of
galaxies along the sightlines toward high-redshift afterglows.  This galaxy 
sample includes both GRB host galaxies and those foreground DLA galaxies that 
are close to the sightlines.  While intervening DLA systems arise 
preferentially in the outskirsts of distant galaxies (because of a gas 
cross-section selection effect), the GRB host sample offers a unique 
opportunity to study ISM physics more immediate to vigorous star-forming 
regions in high-redshift galaxies (that have relatively small cross-section).

  Here we present the first echelle spectrum of a {\it Swift} GRB, obtained 
four hours after the burst.  The spectrum, which spans a wavelength range from
3300 \AA\ through 9400 \AA, exhibits abundant absorption features superimposed 
on a simple power-law shaped continuum.  We identify the GRB at $z_{\rm GRB}=
3.969$ based on a strong damped absorption trough centered at 6040 \AA\ and a 
suite of associated metal absorption features.  In addition, we identify an 
intervening DLA system at $z_{\rm DLA}=3.564$ and a number of strong Mg\,II 
absorbers along the sightline.  We demonstrate that rapid echelle spectroscopy 
is plausible for well-localized afterglows and that the observations provide
new insight into the nature of GRB host environments, as well as the physical
properties of the intergalactic medium (IGM) along the sightlines.

\section{OBSERVATIONS AND DATA REDUCTION}

  We observed the OT of GRB050730 that was first reported by Hollan \etal\ 
(2005) and later confirmed by Sota \etal\ (2005) at RA(J2000) = 
14\fh08\fm17\fs.14 and Dec(J2000)= $-$03\fd46\arcmin17\arcsec8, using the MIKE 
echelle spectrograph (Bernstein \etal\ 2003) on the 6.5 m Magellan Clay 
telescope at Las Campanas Observatory.  The spectrograph contains a blue camera
and a red camera, allowing a full wavelength coverage from near-ultraviolet 
3300 \AA\ through near-infrared 9400 \AA.  The observations were carried out in
a sequence of three exposures of duration 1800 s each, starting at UT 00:00 on 
2005 July 31 (four hours after the initial burst) when the OT had $R\approx 
17.7$ (Holman \etal\ 2005).  The mean seeing condition over the period of 
integration was $0.7''$.  We used a $0.7''$ slit and $2\times 2$ binning 
during readout, yielding a spectral resolution of FWHM $\approx 10$ \kms\ at 
wavelength $\lambda=4500$ \AA\ and $\approx 12$ \kms\ at $\lambda=8000$ \AA.  
The data were processed and reduced using the MIKE data reduction software 
developed by Burles, Prochaska, \& Bernstein\footnote{see 
http://web.mit.edu/\char'176burles/www/MIKE/mike\_cookbook.html.}.  Wavelengths
were calibrated to a ThAr frame obtained immediately after each exposure and
subsequently corrected to vacuum and heliocentric wavelengths.  Flux 
calibration was performed using a sensitivity funtion derived from observations
of the flux standard NGC7293.

  The final stacked spectrum of the afterglow is presented in Figure 1.  The 
signal-to-noise of the echelle data is $S/N=12$ at 4800 \AA\ and $S/N=13$ at 
8000 \AA\ per resolution element.  We have performed a $\chi^2$ fitting 
routine over the spectral region between 7000 \AA\ and 9000 \AA, corresponding
to a wavelength range between 1400 \AA\ and 1800 \AA\ in the rest frame of the
host) and obtained a best-fit power-law model $f_\nu(\lambda)\propto 
\lambda^{\alpha}$ with $\alpha =1.88\pm 0.01$, which is presented as the dotted
curve in Figure 1.  The best-fit power-law index is significantly steeper than 
what has been measured for the OT of GRB020813 ($\alpha=1$) at rest-frame 
optical wavelengths (Barth \etal\ 2003).

\section{ANALYSIS}

  The echelle spectrum of the OT of GRB050730 exhibits a large number of 
absorption features due to the \hI\ \lya\ transition and various heavy ions 
in the ISM of the GRB host as well as in intervening gaseous clouds at 
$z<z_{\rm GRB}$.  Most notably is the strong damping trough centered at roughly
6040 \AA, blueward of which we observe a forest of absorption features that are
not present on the red side of the damped absorption profile.  We identify the 
damped absorber as the DLA originating in the host galaxy of the GRB at $z_{\rm
GRB}=3.969$ and the forest lines as the \lya\ forest at $z<z_{\rm GRB}$.  In 
addition, we also identify an intervening DLA at $z_{\rm DLA}=3.564$ and a 
number of strong metal-line absorbers.  Here we summarize the physical 
properties of these strong absorbers.

\subsection{The GRB Host Environment at $z_{\rm GRB}=3.969$}

  The redshift measured for the GRB host using the DLA feature is confirmed by
associated metal-line transitions.  We measure a more precise redshift of the
GRB at $z_{\rm GRB}=3.96855\pm 0.00005$ using narrower metal absorption lines.
At this redshift, we measure $N(\hI)$ of the GRB host by fitting Voigt profiles
to the observed \lya\ and \lyb\ transitions simultaneously using the VPFIT 
software
package\footnote{see http://www.ast.cam.ac.uk/\char'176rfc/vpfit.html.}.  We
obtain $\log\,N(\hI)=22.15\pm 0.05$ for the host, with the error estimated from
varying the continuum level of the absorption line profiles.  This is the 
highest $N(\hI)$ observed in DLA systems, including those arising in GRB hosts 
(c.f.\ Vreeswijk \etal\ 2004).  

\begin{figure*}
\begin{center}
\includegraphics[scale=0.6, angle=270]{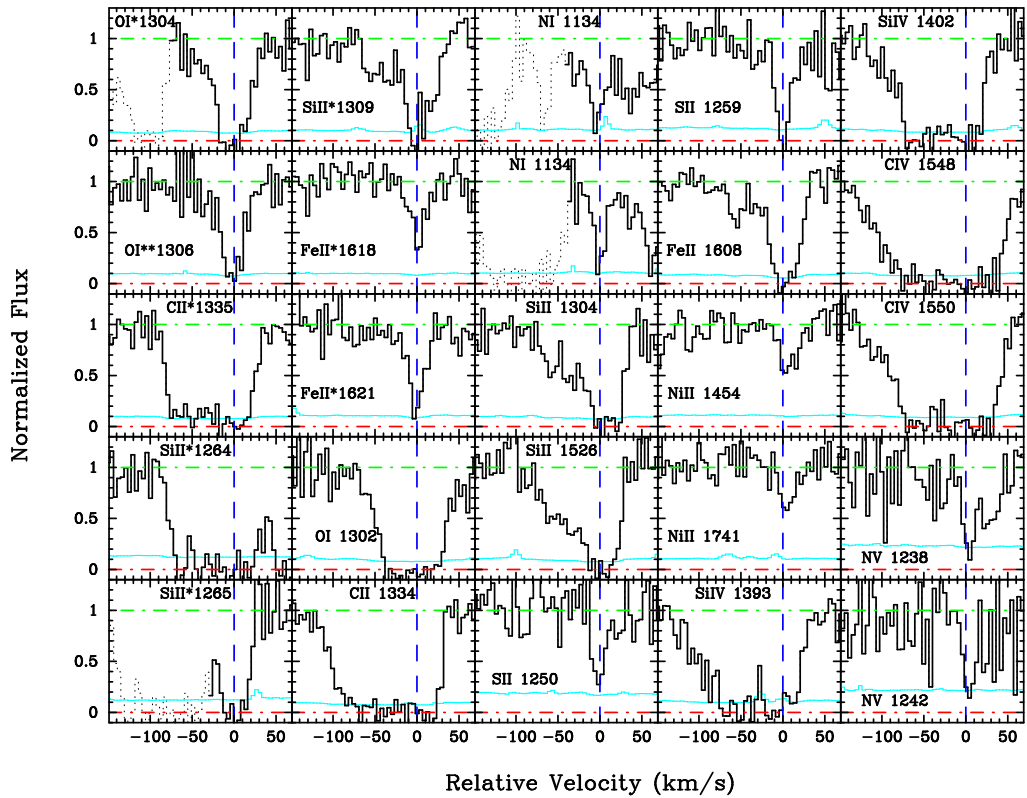}
\caption{Absorption profiles of different ionic transitions found at the
redshift of the GRB host.  The 1-$\sigma$ error spectrum is presented as the 
thin (cyan) curve in each panel.  The zero relative velocity corresponds to 
redshift $z=3.96855$.}
\end{center}
\end{figure*}

  In addition to absorption features due to neutral hydrogen, the spectrum also
exhibits a suite of metal-absorption lines at the GRB host redshift (Figure 2),
from neutral species such as O\,I, to low-ionization transitions such as C\,II,
Si\,II, S\,II, Ni\,II, and Fe\,II, and to high-ionization transitions such as 
C\,IV, Si\,IV, and N\,V.  We also identify strong fine structure lines such as 
O\,I$^*$, O\,I$^{**}$, Si\,II$^*$, C\,II$^*$, and Fe\,II$^*$.  With the 
exception of C\,II$^{*}$ (Wolfe, Prochaska, \& Gawiser 2003), none of these 
transitions has been detected in intervening DLA systems.  In particular,
Fe\,II$^*$ absorption features are only observed locally in Eta Carinae 
(Gull \etal\ 2005).  These saturated fine structure transitions indicate an
extreme ISM environment with high gas density that is rarely observed in
intervening DLA systems.  Furthermore, the profiles of well resolved lines 
(e.g.\ S\,II\,1250) show that $>90$\% of the neutral gas is confined to a 
velocity width 20 \kms, which is considerably smaller than the median value of 
intervening DLA systems and implies a quiescent environment.  But the profiles 
of saturated lines do exhibit absorption extending to $\approx 80$ \kms\ 
(e.g.\ Si\,II 1526).  Finally, the asymmetry of these line profiles is 
suggestive of an organized velocity field, e.g.\ rotation or outflow.

  We measure the column density of each transition using both the apparent 
optical depth method (Savage \& Sembach 1991) and the VPFIT software package.  
Comparing column density ratios between different transitions allow us to 
constrain the physical properties of the ISM in the GRB host.  The principal 
results are as follows.  (1) We measure $[{\rm S}/{\rm H}]=-2.0\pm 0.1$. 
Because S is non-refractory, its gas-phase abundance gives a direct measurement
of the gas metallicity.  (2) We find $[{\rm S}/{\rm Fe}]=+0.3$, consistent with
the gas phase $[\alpha/{\rm Fe}]$ measurements of low-metallicity DLA systems.
Even if we adopt an intrinsic solar abundance pattern, the dust-to-gas ratio in
the host ISM is very low (c.f.\ Savaglio \etal\ 2003).  (3) We measure $[{\rm 
N}/{\rm S}]=-1.0 \pm 0.2$, again consistent with low-metallicity DLA systems 
(Prochaska \etal\ 2002).  (4) We detect no molecular lines in the \lya\ forest,
suggesting a warm gas phase.  (5) Based on the observed ratio of $N({\rm 
Fe\,II}^*)/N({\rm Fe\,II})$, we infer a number density $n_{\rm H} > 10^3\ {\rm 
cm^{-3}}$ for a temperature $T < 30000$ K.  This constrains the size of the 
host DLA cloud to be $l_{\rm DLA} < 4.6$ pc.

  In summary, aside from the large gas density $n_{\rm H}$ inferred from 
saturated fine structure lines, the ISM of the GRB host has very similar 
characteristics to known DLA systems at $z\sim 4$.

\subsection{Intervening Absorbers}

  We have identified a number of strong absorbers along the sightline toward 
GRB050730.  We briefly summarize each system below:

  {\it DLA system at $z_{\rm DLA}=3.56439$}: We identify \lya\ and \lyb\ 
transition for this absorber and measure $\log\,N(\hI)=20.3\pm 0.1$.  In 
addition, we also find associated metal-line absorbers due to Si\,II, Al\,II,
Fe\,II, Si\,IV, and C\,IV.  The weak line strengths of these metal absorption
lines suggest a low metallicity in the neutral gas, $[{\rm Si}/{\rm H}]<-1.3$.

  {\it Lyman Limit system at $z_{\rm LLS}=3.02209$}: We measure $N(\hI)$ using
the \lya\ feature and find $\log\,N(\hI)=19.9\pm 0.1$ for this Lyman limit
system.  In addition, we identify Si\,II, Al\,II and Fe\,II, but we do not 
detect C\,IV absorption at the absorber redshift.  Including no ionization 
correction, we find $[{\rm Si}/{\rm H}]=-1.5\pm 0.2$.  

  {\it Mg\,II system at $z_{\rm Mg\,II}=2.25313$}: We identify saturated Mg\,II
doublet at this redshift.  In addition, we find strong Mg\,I\,2852 and Fe\,II 
absorption features for this absorber.

  {\it Double Mg\,II systems at $z_{\rm Mg\,II}=1.7731$}: We identify a pair of
Mg\,II absorbers at $\Delta v=57$ \kms\ apart, for which we also find 
Mg\,I\,2852 and the Fe\,II absorption series.

\section{DISCUSSION AND CONCLUSIONS}
 
  GRB afterglows clearly provide a novel alternative to quasars for probing the
ISM and IGM in the young universe.  In the case of GRB050730, the afterglow 
reached an initial brightness of $R=15.5$ (Klotz \etal\ 2005) and faded to 
$R=17.5$ four hours later when the echelle spectroscopy was commenced.  
At $z=3.969$, $R=15.5$ corresponds to an absolute magnitude $M_{1450}=-29.3 + 
5\log h$ at rest-frame 1450 \AA\footnote{Adopting a $\Lambda$ cosmology, 
$\Omega_{\rm M}=0.3$ and $\Omega_\Lambda = 0.7$, with a dimensionless Hubble 
constant $h = H_0/(100\ {\rm km} \ {\rm s}^{-1}\ {\rm Mpc}^{-1})$}.  This
intrinsic luminosity, within a minute after the burst, is comparable to the 
most luminous quasar known (c.f.\ HS 1700$+$6416 at $z = 2.73$; Schneider 
\etal\ 2003).  Even at $R=17.5$, the corresponding intrinsic luminosity 
easily competes with the brightest quasars known at this redshift (see e.g.\ 
Fan \etal\ 2001).  While absorption line systems uncovered toward the 
sightlines of optically selected quasars are likely to miss dusty absorbers, 
the extreme brightness of early GRB afterlows offers an unbiased view of the 
IGM at all epochs.  In addition, the simple power-law shaped continuum of a GRB
afterglow allows a more precise and accurate measurement of IGM opacity.

  The sightline toward GRB050730 is particularly interesting with a strong DLA 
feature arising in the GRB host galaxy and an intervening DLA system at lower 
redshift.  The GRB host DLA system ($z_{\rm DLA}=z_{\rm GRB}$), which are
presumably selected by vigorous star formation and therefore probe deep into 
the center regions of distant galaxies, presents a nice contrast to the 
intervening DLA system ($z_{\rm DLA}<z_{\rm GRB}$), which arises preferencially
at large galactocentric radii due to a larger cross-section of the outskirts
than the inner regions.  Our study shows that aside from having a much higher 
neutral gas density, the GRB host DLA has very similar characteristics to known
$z\sim 4$ DLA systems, such as low dust content, low metallicity, and 
$\alpha$-element enhanced chemical composition.

  We have also shown that this sightline runs through a number of strong 
intervening absorbers, including a Lyman limit system and two strong Mg\,II
absorbers.  Detailed analyses of various ionic transitions associated with
these absorbers allows us to study physical properties of ISM and IGM over a 
wide redshift range from $z=1.7$ through $z=3.9$.  Follow-up deep imaging and 
low-resolution spectroscopy of faint galaxies along the line of sight will 
allow a direct comparison between the physical properties of the cold ISM (such
as metallicity, kinematics, and dust content as derived from absorption line 
studies) and stellar properties (such as luminosity, morphology, and star 
formation rate as extracted from absorbing-galaxy analyses).  A large sample of
GRB host DLA systems offers a unique opportunity to dissect the ISM properties 
that are directly connected to GRBs, while a statistical sample of intervening 
DLA systems identified toward GRB sightlines will provide important insights 
toward understanding the nature of high-redshift DLA systems and offer a simple
test to discriminate between different galaxy formation scenarios (e.g.\ 
Haehnelt, Steinmetz, \& Rauch 2000).

\acknowledgments
 
  We appreciate the expert assistance from the staff of the Las Campanas 
Observatory.  It is a pleasure to thank John O'Meara and Scott Burles for 
assistance with the data reduction, and Chris Howk and Art Wolfe for helpful 
discussions.  H.-W.C., JXP, and JSB acknowledge support from NASA grant 
NNG05GF55G.



\begin{references}

\vskip 0.2in

\reference{} Akerlof, C. \etal\ 1999, Nature, 398, 400

\reference{} Barth, A.~J. \etal\ 2003, ApJ, 584, L47

\reference{} Bernstein, R. \etal\ 2003, Proc. SPIE, 4841, 1694

\reference{} Blain, A.~W. \etal\ 2002, Phys. Rep., 369, 111

\reference{} Bloom, J.~S. \etal\ 2002, ApJ, 572, L45

\reference{} Bloom, J.~S. \etal\ 2005, ApJ in press (astro-ph/0505480)

\reference{} Castro, S. \etal\ 2003, ApJ, 586, 128

\reference{} Chen, H.-W. \& Lanzetta, K. M. 2003, ApJ, 597, 706

\reference{} Colbert, J. W. \& Malkan, M. A. 2002, ApJ, 566, 51

\reference{} Fan, X. \etal\ 2001, AJ, 121, 54

\reference{} Fiore, F. \etal\ 2005, ApJ, 624, 853

\reference{} Gull, T.~R.\ 2005, ApJ, 620, 442

\reference{} Haehnelt, M.~G., Steinmetz, M., \& Rauch, M. 2000, ApJ, 534, 594

\reference{} Holland, S.~T. \etal\ 2005, GCN Circ. 3704 
             (http://gcn.gsfc.nasa.gov/gcn3/3704.gcn3)

\reference{} Holman, M., Garnavich, P., \& Stanek, K.~Z. 2005, GCN Circ. 3727 
             (http://gcn.gsfc.nasa.gov/gcn3/3727.gcn3)

\reference{} Jakobsson, P.\ 2004, A\&A, 427, 785

\reference{} Jensen, B.~L.\ 2001, A\&A, 370, 909

\reference{} Klotz A., Boer M., \& Atteia J.~L. 2005, GCN Circ. 3720
             (http://gcn.gsfc.nasa.gov/gcn3/3720.gcn3)

\reference{} Le Brun, F. \etal\ 1997, A\&A, 321, 733

\reference{} M{\o}ller, P.\ \etal\ 2002, A\&A, 396, L21

\reference{} Paczy\'nski, B. 1998, ApJ, 494, 45

\reference{} Pettini, M. \etal\ 1999, ApJ, 510, 576

\reference{} Prochaska, J.~X. \etal\ 2002, PASP, 114, 933

\reference{} Prochaska, J.~X. \etal\ 2003, ApJS, 147, 227

\reference{} Prochaska, J.~X. \etal\ 2004, ApJ, 611, 200

\reference{} Rao, S.~M.\ \etal\ 2003, ApJ, 595, 94

\reference{} Savage, B.~D. \& Sembach, K.~R. 1991, 379, 245

\reference{} Savaglio, S., Fall, M.~S., \& Fiore, F.\ 2003, ApJ, 585, 638

\reference{} Schneider, D.~P. 2003, AJ, 126, 2579

\reference{} Sota, A. \etal\ 2005, GCN Circ. 3705
             (http://gcn.gsfc.nasa.gov/gcn3/3705.gcn3)

\reference{} Stanek, K.~Z. \etal\ 2003, ApJ, 591, L17

\reference{} Steidel, C. C. \etal\ 1999, ApJ, 519, 1

\reference{} Vreeswijk, P. \etal\ 2004, A\&A, 419, 927


\reference{} Wolfe, A.~M., Prochaska, J.~X., \& Gawiser, E., 2003, ApJ, 593, 
             215

\reference{} Wolfe, A.~M. \etal\ 2005, ARA\&A, 43, 200

\reference{} Woosley, S.~E. 1993, ApJ, 405, 273

\end{references}
\end{document}